\documentclass[aps,twocolumn,groupedaddress, showpacs,prb]{revtex4}
\usepackage{graphicx}%

\begin{document}
\bibliographystyle{apsrev}


\title{A graphite-prism definition for Avogadro's ``integer"}


\author{P. Fraundorf and Melanie Lipp}
\affiliation{Physics \& Astronomy/Center for Nanoscience, U. Missouri-StL (63121) USA}
\email[]{pfraundorf@umsl.edu}


\date{\today}

\begin{abstract}

The new International System of Units may let us select an integer value for Avogadro's number. Some might prefer an integer that's divisible by 12, so that an integer number of $^{12}C$ atoms may be associated (to first order) with a gram's mass. For educational and practical reasons it may also help to choose a {\em physically-meaningful} definition within measurement error of the current numeric value. Cubes of diamond face-centered-cubic Si and (much rarer) face-centered-cubic C have been proposed, but these structures do not have naturally-occurring facets (or numbers of atoms generally divisible by 12). We show here that graphite prisms formed by stacking $m$ hexagonal graphene sheets, with $m \equiv 51,150,060$ carbon-12 atoms on each side, are a natural solution that may facilitate generation of precise molar standards as well.  

\end{abstract}
\pacs{81.05.uf, 06.20.fa, 06.20.Jr, 61.48.Gh}
\maketitle

\tableofcontents

\section{Introduction}
\label{sec:Intro}

The upcoming revision of the international system of units\cite{CGPM2014} will give the Avogadro constant $N_A$ an exact numerical value, both as a mass-independent number of elementary entities and {\em possibly}\cite{Leonard2012} as the number of grams/Dalton regardless of the precise number of Daltons associated with any elementary entity. Thus the mass $M$($^{12}C$) of a $^{12}C$ atom may become an experimental quantity (as always affected by neighbor binding interactions) only approximately equal to 12 Daltons\cite{Mills2006, Hill2011, Wheatley2011a, Milton2013}. Since the numeric value of the Avogadro constant will be chosen for consistency with existing values e.g. to 8 significant figures, this paper discusses physically-meaningful choices for the other 16 (or more) significant figures.

Most likely agree that if we get to select Avogadro's constant as some number of elementary entities (e.g. atoms or molecules) that it might as well be an integer. For pedagogical purposes in discussing units for mass, it might also help\cite{Khruschov2010} to make that integer divisible by 12 e.g. so that one gram is also to first order the mass of an integer number of $^{12}C$ atoms.

One strategy for doing this might be to require as few significant decimal digits as possible to specify the constant, since most humans use base-10 numbers. For example $602,214,150,000,000,000,000,000$ is an integer divisible by 12 that would serve admirably.

The other strategy is to choose the number of atoms in a relevant physical structure, to make the definition concrete. Cubes of simple-cubic and face-centered-cubic carbon\cite{Fox2007}, and of diamond face-centered-cubic silicon\cite{Khruschov2010}, have for example been proposed in this context.

The major problems with these choices are that: (i) simple-cubic carbon doesn't exist in nature, (ii) fcc carbon is at best rare\cite{Konyashin2006,Tapia2005}, (iii) none of these structures have naturally-occurring facets, during growth or cleavage, that lie on the (100) planes which bound these cubes, and (iv) symmetry does not dictate that the number of atoms in these structures is divisible by 12.

We show here that a good approximation (divisible by 12) is provided by hexagonal graphite prisms with $m$ graphene-layers having m-atom armchair edges where $m=51,150,060$. Graphite in turn is constructed from graphene sheets whose controlled synthesis at the atomic-scale is likely to see great progress by nanotechnologists in the years ahead. The atoms in an individual graphene sheet can already be counted\cite{Krivanek2010}, and by extension these structures in tube form may downstream allow one to generate macroscopic molar-standards whose accuracy is limited primarily by one's ability to cut off a well-defined length. 

\section{Growing graphene}
\label{sec:growing}

\begin{figure}
\includegraphics[scale=0.70]{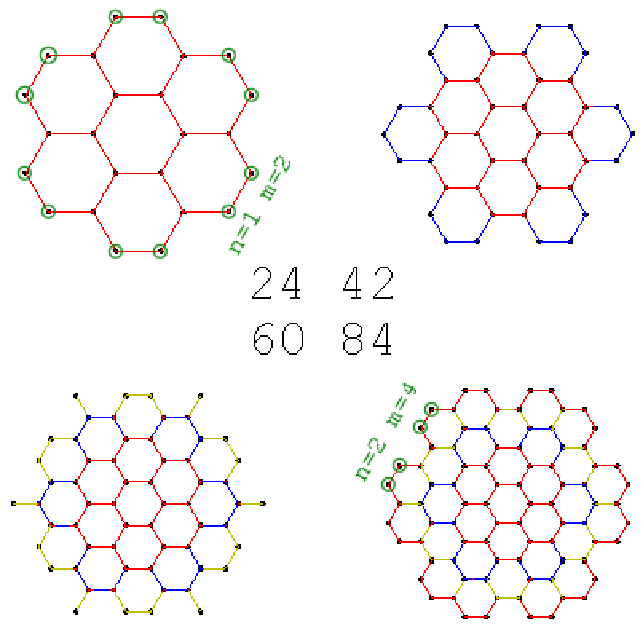}%
\caption{Graphene hexagons of increasing size.}
\label{fig1}
\end{figure}

Our examination of graphene structures begins with a Mathematica program designed to add ``melt-carbon" atoms to sheets with defects, to examine dendritic crystallization of unlayered-graphene from a non-equilibrium melt as one possible mechanism to explain the unlayered-graphene cores\cite{Bernatowicz1996,Fraundorf2002,Mandell2007} found in a subset of pre-solar graphite onions. Although the program is designed to add carbon atoms to sheets with defects, here we've simply added carbon atoms to a flat "graphene seed". 

In Fig. \ref{fig1} we start with 24 atoms bounded by a pair of adjacent ``armchair-edge" atoms on each of the 6 sheet-sides for $2\times 6 = 12$ surface atoms with only two bonds and 12 internal atoms with 3 bonds each. This ``seed-state" has $6 + 6 + 12 = 24$ atoms, and can be used as a starting point for ``zigzag edge" hexagons as well.

The first growth step (top left to top right) adds three external edge or ``sheet-surface" atoms (with fewer than 3 ``bonded" neighbors) between adjacent ``arm-chair atom-pairs", leaving us now with 24 internal atoms and $ 3 \times 6 = 18$ surface atoms for 42 atoms total. The sequence of additions now looks like $6 + 6 + 12 + 18 = 42$.

The second growth step (top right to bottom left) adds a singly-bonded edge atom and two doubly-bonded edge atoms to each of the previous 6 hexagon sides. Now there are 42 triply-bonded internal atoms, $6 \times  2 = 12$ doubly-bonded surface atoms and $6 \times  1 = 6$ singly-bonded surface atoms for a total of 60 atoms. The sequence here is $6 + 6 + 12 + 18 + 18 = 60$.

Finally we add 4 doubly-bonded edge-atoms on each side. Now there are 60 triply-bonded internal atoms, and $6 \times 4 = 24$ doubly-bonded surface atoms with a total of 84 atoms in the same closed-shell arrangement of the starting sheet. The sequence is now $6 + 6 + 12 + 18 + 18 + 24 = 84$.

One possible pattern of atom-increments is $(0) + 6 + 6 + (12) + 18 + 18 + (24) + 30 + 30 + (36) + 42 + 42 + (48)$ etc. The totals would then read $(0), 6, 12, (24), 42, 60, (84), 114, 144, (180), 222, 264, (312)$ etc. Jumping only between the ``closed-shells'' (with dips between pairs at each corner) in parentheses, the total number of carbon atoms goes through $0, 24, 84, 180, 312, 480$ etc. with $0, 6, 12, 24, 36, 48, 60$ etc. edge-atoms respectively.

The closed-shell recurrence relation for the total number N of atoms therefore looks like $N_{n+1} = N_n + 3S_n + 24$, where the number of edge-atoms $S_{n+1} = S_n + 12$. From this, it looks like closed shells with n arm-chair atom-pairs along each of 6 sides (i.e. 2n ``edge-atoms") contain $N_{hex} = 6n(1+3n) = 3 m + 9 m^2/2$ carbon atoms with $S_{hex} = 12n = 6m$ atoms on the hexagonal sheet edge, in terms of the number of atom-pairs $n = 0, 1, 2,$ etc. and the number of atoms $m \equiv 2n = 0, 2, 4,$ etc. on each side. 

\begin{figure}
\includegraphics[scale=0.75]{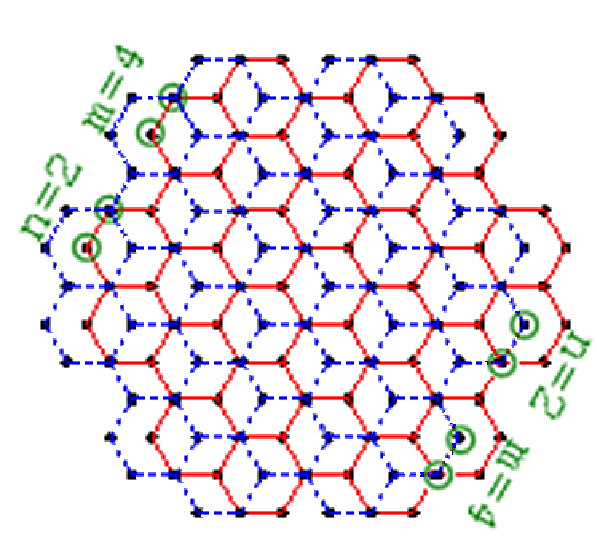}%
\caption{Stacking graphene $\Rightarrow$ graphite.}
\label{fig2}
\end{figure}

\section{Stacking hexagons}
\label{sec:stacking}

Now consider an even-integer number of $m$ hexagonal graphene-sheets in a graphite stack, such that $m$ atoms on each hexagonal sheet-edge yields a total number of carbon atoms closest to our standard value for Avogadro's number. The even-integer choice of $m$ for atoms on the side of a closed-shell sheet comes from their natural pairing along the sheet edges, while the even-integer choice of $m$ for number of sheets arises from symmetry considerations because the Bernal-structure\cite{Bernal1924} ABAB unit cell\cite{Charlier1992} is two-layers thick in the (002) direction as shown in Fig. \ref{fig2}.

\begin{figure}
\includegraphics[scale=0.35]{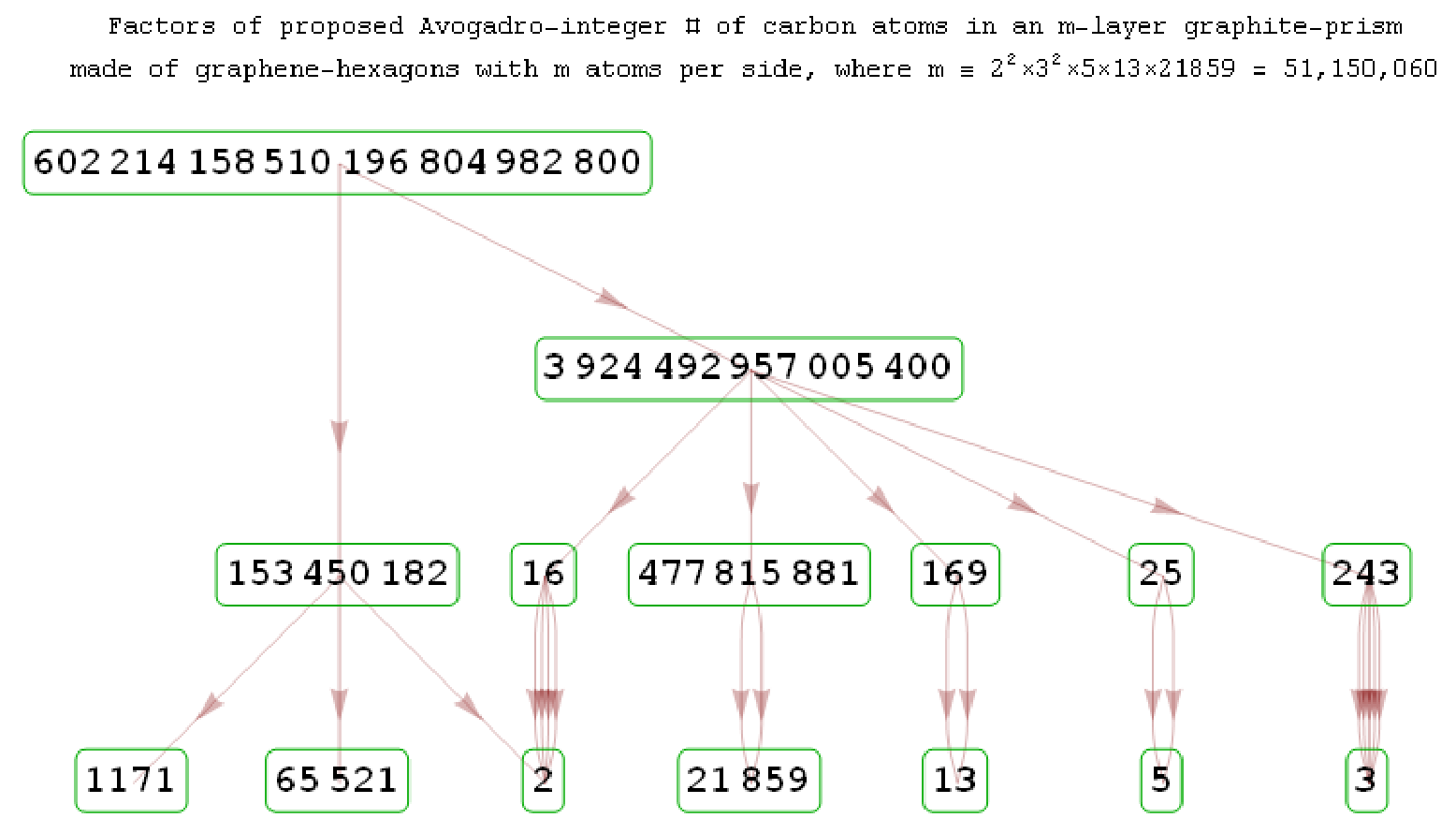}%
\caption{Possible factors for Avogadro's number as a physically-realizable number of carbon atoms, divisible by 12.}
\label{fig3}
\end{figure}

The number of atoms in a hexagonal prism constructed in this way, with an even number of $m$ hexagonal sheets with $m$-atoms per armchair edge, is:
\begin{equation}
N_{total} = m N_{hex} = 3m^2+9m^3/2.
\label{choice}
\end{equation} 
If $m=0$, the $N_{total} = 0$. If m = 2, $N_{total} = 2 \times 24 = 48$ atoms. If m = 4, $N_{total} = 4 \times 84 = 336$ atoms, etc.

Although not obvious at first glance, $N_{total}$ will always be divisible by 12. One can see this from our physical model by pointing out that any structure with even $m$ can separated into two equal layer-sets, each of which in turn (by symmetry) can be divided into six equal pie-slices. Mathematically, this is true since $N_{total}/12 = (m/2)^2(1+3(m/2))$ is an integer if $m$ is even.

Of course we are looking for $m$ such that $N_{total}$ becomes Avogadro's number. For example a faceted graphite-crystal with a set of $m$ (002) graphene sheets, each of which has $m$ atoms along the sheet hexagonal edges that make up the 6 \{110\} facets, would for $m = 51,150,060$ have $602,214,158,510,196,804,982,800$ atoms which compares nicely to current approximations.

The result of this choice for m is closer to the current experimental value $6.0221415 \times 10^{23}$ than is the best (but non-realizable simple-cubic) model proposed in the Fox article about carbon cubes\cite{Fox2007}. It is also closest of those proposed models to the value of $6.02214179 \times 10^{23}$ recommended here\cite{Mills2010}, and subtracting 2 from this value of m will put it closer to the lower value of $6.02214078 \times 10^{23}$ based on Si-28 measurements here\cite{Andreas2011}. 

The coincidental nature of this result is emphasized by the fact that the best multiple-of-12 match to Avogadro's number with a zigzag-edge graphite prism (with $m=46,472,916$) is only accurate to 6 instead of 8 decimal points. Literature on monolayer and folded sheets suggest that both types of edges occur\cite{Koskinen2008}, although armchair may be slightly more abundant\cite{Liu2009}. Bilayer edges appear even more stable\cite{Qi2010}, although ways to choose armchair over zigzag edges on macroscopic prisms (e.g. grown from a Ni melt) remain unclear at this point\cite{Amini2013}. Edge and corner reconstructions might further complicate the maintenance of structure at the atomic scale, on a ``mole-sized" standard in particular.

The resulting armchair-graphite hex-prism would be about 1.71 cm thick, and have a (circumscribed-cylinder) diameter (twice the length of one side) of about 2.18 cm. A hexagonal prism like this is already approximately one-mole of carbon. If the physical model discussed here is used to redefine Avogadro's number, to the extent that Carbon-12 has a mass at or near 12 Daltons then one gram would be near if not equal to the mass of $N_A /{12} = 50,184,513,209,183,067,081,900$ atoms of $^{12}C$ as well. 

The divisibility of this suggested integer is related to its high factorability. As shown in Figure \ref{fig3}, the factors include five factors of 2 and five factors of 3. The hex-prism layer/side integer $m = 51,150,060$ is also relatively factorable, with two factors of 2, two of 3, and one each of 5, 13 and 21859.

\section{Discussion}
\label{sec:discuss}

\begin{figure}
\includegraphics[scale=0.25]{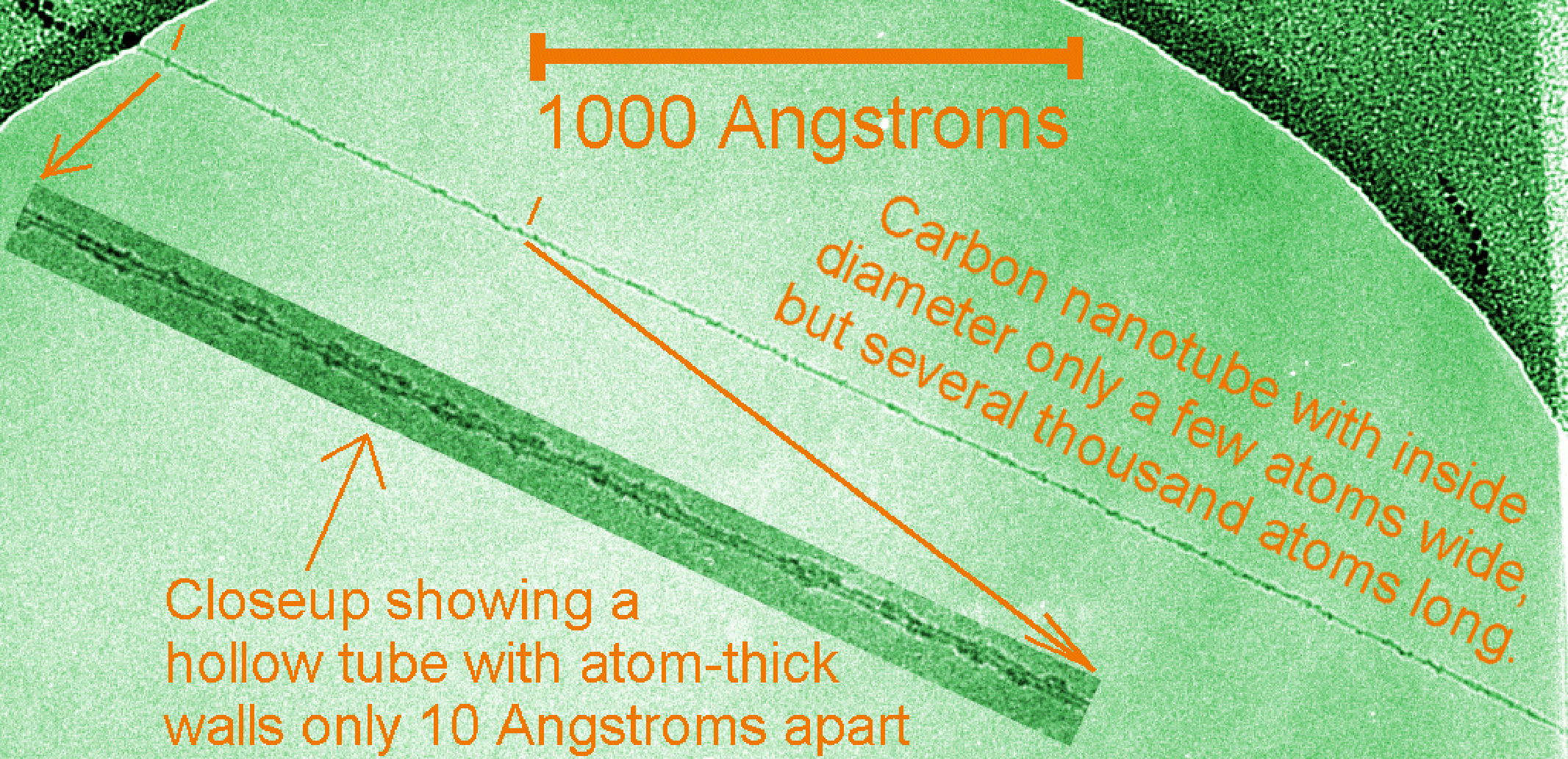}%
\caption{Electron phase-contrast image of a single-walled carbon 
nanotube\cite{Unrau2009} not much larger in diameter than a buckyball.}
\label{fig4a}
\end{figure}

The Avogadro constant is in essence a conversion factor between atom-scale and macroscopic units. Its scale size is therefore specified to many significant figures historically, by the relationship between macroscopic and atomic units of mass. If the new international system allows us to select an exact value for this constant, there may be scientific and pedagogical value for using a physically-meaningful model.

The scientific value, for example, might be found in our ability to construct and maintain precise molar/mass standards by simply counting the number of atoms (e.g. with an electron microscope) to make sure all atoms are still there before putting the standard to use. Carbon and its graphite/graphene structures are already among the most intensely-studied by nano-technologists, and the most stable at least in the absence of hot oxygen or molten iron. 

Thus a machine able to ``turn-out" a nanotube of arbitrary length might be able to dispense molar quantities precisely defined by that length, since for instance a one-meter length of the 1 nm diameter tube shown in the experimental image (Fig. \ref{fig4a}) would contain only $\simeq 200$ femto-moles of carbon. Note that the experimental uncertainty associated with exact molar quantities in the new SI will reside in the component masses, including $M$($^{12}C$), and interaction binding energies rather than in the molecule to mole conversion itself as is the case today. 

In fact, specimens small enough\cite{Treacy1996,Krivanek2010} in which to ``count and measure the positions of carbon atoms" might someday facilitate improved experimental determinations of $M$($^{12}C$) in Daltons (and hence the number of grams per mole). The new-SI distinctions between number of atoms and specimen mass might be helpful for experiments with an interest in sub-ppb accuracy, like this, because mass-deficits associated with chemical-bond geometry become significant below the ppb level since particles add simply but masses do not.

\begin{figure}
\includegraphics[scale=0.50]{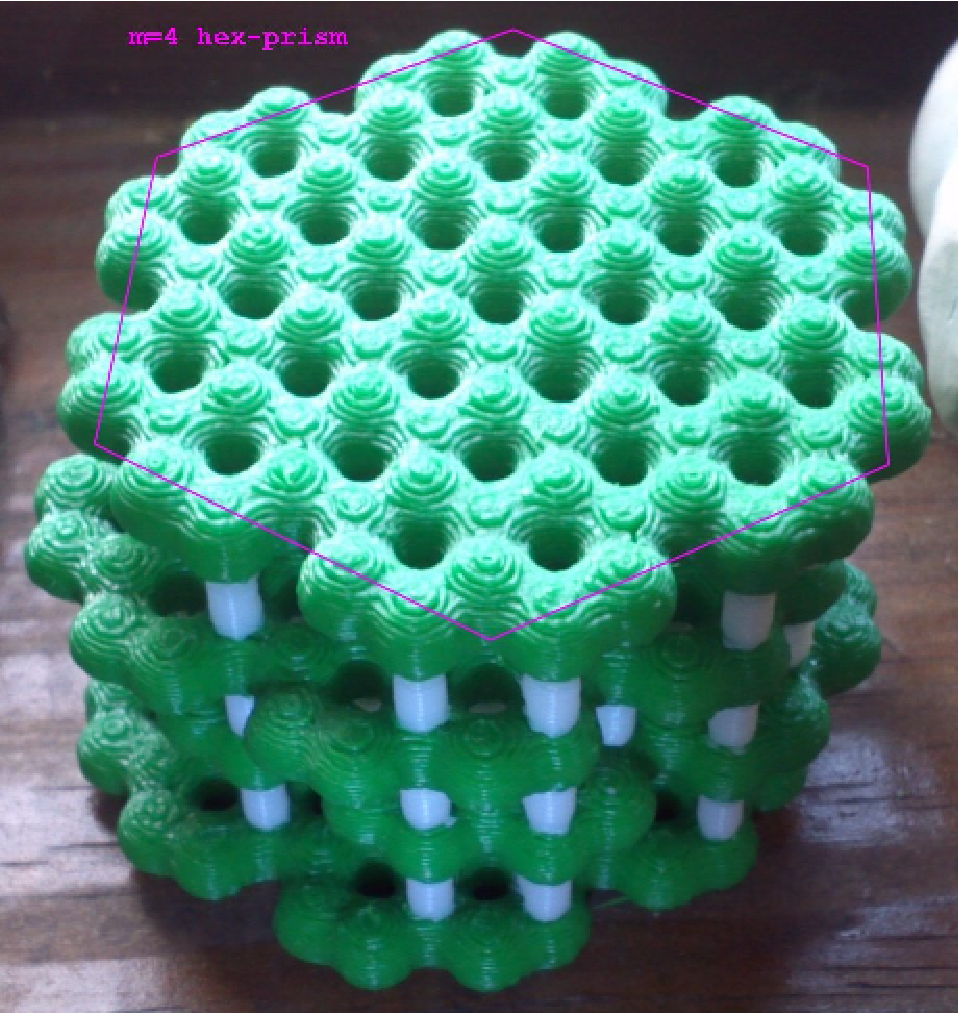}%
\caption{Three-dimensional print of a graphite $m=4$ hex-prism at 100pm to 3mm scale, with 13 pm height-contours.}
\label{fig4}
\end{figure}

Independent of our ability to manufacture and maintain 12-gram (or much smaller) graphite standards which contain a precise number of moles of $^{12}C$, on the pedagogical front one can certainly build graphite and ``pretend-graphite" models of the hexagonal prism. Figure \ref{fig4} illustrates an $m = 4$ hex-prism, with a form-factor similar to that of the $m = 51,150,060$ version. This can be constructed in the rescaled size of the Avogadro's number prism using today's 3D printing technology.

Boltzmann's constant in the new-SI might similarly be defined as a conversion factor\cite{pf.hcapbit} using $d(S/k_B)/dE=1/(k_B T)$ to define $1$ [J/K] as $1/k_B$ [nats] $\equiv 1/(k_B \ln[2])$ [bits] of uncertainty. A target $k_B$ value (to arbitrary pecision) might thereby be used to define the J/K ``information unit'' in [bits] as the product of a binary multiple (e.g. $2^{61}$) times a prime number (e.g. $45,317$).

\begin{acknowledgments}
Co-author PF stumbled upon the coincidental match
between this structure and present values for 
Avogadro's number, while co-author ML has been 
instrumental in our calculation of zig-zag versus 
arm-chair edge-energies as well as in clarifying the 
presentation.
\end{acknowledgments}



\bibliography{ifzx2}

\end{document}